\documentclass[aps,prb,showkeys,preprint,showpacs,12pt]{revtex4-1}
\usepackage{amssymb}
\usepackage{hyperref}
\usepackage{graphics,graphicx}
\usepackage{dcolumn}
\usepackage{amsbsy}
\usepackage{amsmath}
\usepackage{epsfig}
\usepackage{color}
\usepackage{textcomp}
\begin{document}
 \title{Observation of spin-glass behavior in nickel adsorbed few layer graphene}
\author{Sreemanta Mitra$^{1,2}$}
\email[]{sreemanta85@gmail.com}
\author{Oindrila Mondal$^{3}$}
\author{Sourish Banerjee$^{2}$}
\author{Dipankar Chakravorty$^{1,\dag}$}
\email[]{mlsdc@iacs.res.in}
\affiliation{
$^{1}$
 MLS Prof.of Physics' Unit,Indian Association for the Cultivation of Science, Kolkata-700032, India.\\ }
\affiliation{
$^{2}$
Department of Physics, University of Calcutta, Kolkata-700009, India.\\}
\affiliation{
$^{3}$
Department of Physics, M.U.C. Woman's College, Burdwan, India.\\}

\begin{abstract}
 Nickel-adsorbed graphene was prepared by first synthesizing graphite oxide (GO) by modified Hummers' method and then reducing a solution containing both GO and $Ni^{2+}$. 
EDX analysis showed 31 atomic percent nickel was present. Magnetization measurements under both dc and ac magnetic fields were carried out in the temperature range 
2 K to 300 K. The zero field cooled and field cooled magnetization data showed a pronounced irreversibility at a temperature around 20 K. The analysis of the ac susceptibility 
data were carried out by both Vogel-Fulcher as well as power law.
From dynamic scaling analysis the microscopic flipping time $\tau_{0}\sim 10^{-13} s$ and critical exponent $z\nu=5.9\pm0.1$ were found, 
indicating presence of conventional spin glass in the system. The spin glass transition temperature was estimated as 19.5 K. 
 Decay of thermoremanent magnetization (TRM) was explained by stretched exponential function with a value of the exponent as 0.6 .
 From the results it is concluded that nickel adsorbed graphene behaves like a spin-glass.

\end{abstract}
\maketitle
\section{Introduction}\label{sec.1}
Discovery of graphene has opened up many avenues of research because of basic physics and limitless possibilities of developing novel devices
in the nanoscale,using this material.\cite {geimnat,geimsc} Not only are the electronic properties of unusual nature,\cite{becerrilacsnano,novonat}
the mechanical and thermal behaviour have also been found to be remarkable,the breaking strength and thermal conductivity exhibiting record
values.\cite{leesc,navarronl,balannl} 
Magnetism in graphene due to the presence of defects has been studied theoretically \cite{jpsj76} and  recently observed
experimentally.\cite{wangnl} The edge state induced magnetism in graphene has also been observed both theoretically\cite{jpsj77} and experimentally.\cite{adv22}
 Substitutional nickel impurities have been demonstrated to be present 
in graphenic carbon nanostructures, prepared by using nickel containing catalyst.\cite{ushiroprb,banhartprl} Calculations based on spin
density functional theory have indicated that substitutional nickel defects in flat graphene are non magnetic.\cite{santosprb} A nonzero spin 
moment can however, be observed if the adsorbed nickel atoms lie along the edges of the graphene.\cite{rigoprb} We have synthesized nickel adsorbed graphene, 
starting from graphite oxide using a solution route. Detailed magnetization measurements under both dc and ac magnetic field were carried out. A spin-glass behaviour was observed.
The details are reported in this article.
 \section{Synthesis and Characterization}\label{sec:2}
Primarily, the graphite oxide (GO) was prepared from extra pure fine graphite powder (as obtained from LOBACHEMIE), using modified Hummers' method.\cite{hummersjacs,zhoujpcc}
The graphene oxide was prepared by stirring  powdered flake graphite and sodium nitrate ($NaNO_{3}$), 2 g each and 6 g of potassium permanganate ($KMnO_{4}$) 
($98.5{\%}$ pure, E Merck) into 50 ml concentrated sulfuric acid ($98{\%}$, E-Merck). The ingredients were mixed in a beaker, that had been cooled 
to 273 K in an ice bath. The bath was then removed and the suspension brought to room temperature (300 K) and put under constant 
stirring for 2 h. As the reaction progressed, the mixture gradually thickened with a diminishing in effervescence. After 2 h, 300 ml water and 5 ml 
hydrogen peroxide($H_{2}O_{2}$) were slowly added under stirring. The suspension was then filtered resulting in a yellow-brown filter cake. The filtrate
was washed several times with 1:10 HCl solution in order to remove unwanted metal ions present.\cite{kovtycm} The collected washed sample was dried in an air oven at 333 K
for 2 days. 
\par
Uniform aqueous dispersion of GO (0.01 g in 10 mL) was mixed with 10 mL 0.03 M solution of nickel nitrate (hexahydrate)($Ni(NO_{3})_{2},6H_{2}O$) (E merck, Germany).
The mixture was stirred thoroughly for 2 h. By adding ammonia solution to the mixture its pH was brought to a value $\approx 10$. 0.142 g sodium borohydride ($NaBH_{4}$)
(0.37 M) was added to reduce GO and $Ni^{2+}$ simultaneously to form Nickel adsorbed graphene. The sample was collected by centrifuging the
aqueous mixture at 2000 rpm in an ultracentrifuge (elTek TC 4100D). The sample was then washed thoroughly with deionized water several times to remove any 
unreacted ions present. The sample was then dried at 348 K for 24 h.
\par
The X-ray diffractogram was taken in a Bruker D8 SWAX diffractometer using Cu $K_{\alpha}$ monochromatic source of wavelength ($\lambda=0.154$ nm ). 
The microstructure was studied by using 
transmission electron microscope (JEOL 2010) operated at 200 kV.
To study the chemical composition of the sample,  Energy dispersive X-ray spectroscopy (EDX) 
was done using a JEOL JSM-6700F field-emission scanning electron microscope. Fourier transform infrared spectroscopy (FTIR) studies of the samples were carried out
using FTIR8400S spectrometer.
Both the dc and ac magnetic measurements were carried out by using a Superconducting Quantum Interference Device (SQUID) magnetometer (Quantum Design MPMS XL) 
with the reciprocating sample option (RSO) and a sensitivity of $10^{-7}$ emu. In order to obtain low magnetic fields the SQUID was demagnetized before the 
measurements.
\section{Results and Discussions}\label{sec:3}
  \subsection{Structural Analysis}
    Figure 1(a) shows the x-ray diffractogram obtained from the nickel treated graphene. Only a hump around $25.6^{o}$ is observed due to graphene.\cite{gracnr} 
Since a monolayer graphene is not expected to show any diffraction peak, the small hump signifies that the system comprises few layer graphene. 
No signature of any diffraction peaks corresponding to either nickel or any of its oxides have been observed. The inset of 
figure 1(a) is the transmission electron micrograph of the nickel treated graphene. There is no evidence of any nanocluster being present in the micrograph. 
This is corroborated by the Figure 1(b) which shows the high resolution transmission electron micrograph (HRTEM) of the sample. The only interplanar spacing
 seen, has a value 0.3 nm which corresponds to the (002) reflection in graphene.\cite{gracnr} 
In the inset of figure 1(b) the electron diffraction pattern (SAED) obtained from figure 1(b) is shown. 
The absence of any diffraction spot also rules out the possibility of the presence of 
either Ni or NiO or $Ni_{3}C$ in our system. The weak diffraction ring corresponds to plane (002) of graphene derived from GO.
The absence of any other lattice plane means total absence of any nanoparticle in the system.
 It has, however, not been possible to characterize 
the position of the Ni atoms in graphene. Scanning tunneling microscopy (STM) could not be used because the sample is in powder form and therefore, it was difficult 
to mount in STM and pick up its position with reference to the graphene sheet.
However, on the basis of the computational work reported in the literature \cite{rigoprb,jpcm20} we believe the nickel atoms to be forming bridgelike structure between
two surface carbon atoms in graphene as shown schematically in figure 2. This is the equilibrium structure for the adatoms in graphene as shown theoretically.\cite{prl91}
It is emphasized that these nickel atoms will have magnetic moments as opposed to the situation in which nickel atoms substitute carbon atoms in graphene layer.\cite{santosprb}
\begin{figure}[h]
\centering

\includegraphics[width=8.5cm]{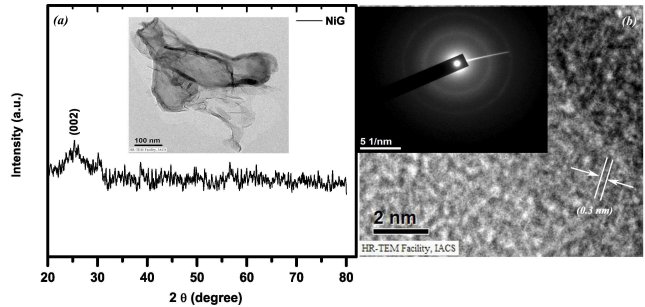}

\caption{(a)X-ray diffractrogram of nickel treated graphene(b) Transmission electron micrograph of NIG.
(c)SAED pattern obtained from (b). (d)HRTEM image of the sample.}
\label{fig.1}
   \end{figure}
\begin{figure}
\centering
\includegraphics[width=8.5cm]{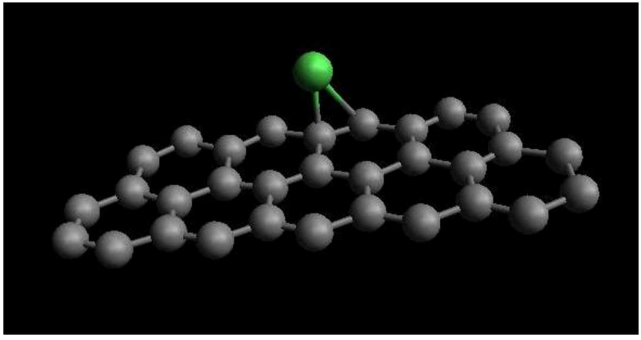}

\caption{Schematic representation of the nickel (green colored atom) adsorded graphene system on the basis of computational work \cite{rigoprb,jpcm20}.}
\label{fig.2}
   \end{figure}

 \subsection{Compositional Analysis}
   The FTIR spectra of graphite oxide (GO), chemically converted graphene (CCG) and nickel adsorbed graphene (NIG) have been shown in figure 3(a).
The treatment of GO with $NaBH_{4}$ causes an enormous structural change (product CCG),which have been observed in FTIR spectrum. 
The spectrum for GO shows transmittance dips at 1390.65, 1634.23, and 1725.14 $cm^{-1}$, which correspond to deformation of O-H bond in water, stretching
mode of carbon carbon double bond (C=C), and stretching mode of carbon oxygen double bond (C=O), respectively. The broad dip at around 3400 $cm^{-1}$ arises 
due to the bending of O-H bond in water\cite{gracnr}.It has been seen that 
there is no transmittance dip at around 1725 $cm^{-1}$,for CCG and NIG. So it is concluded that both CCG and NIG are free from oxygen functionality present in GO.
Another extra transmittance dip (in comparison to CCG ) around 2070 $cm^{-1}$ (marked by red arrow) is observed for the NIG sample, which occurs due to 
the stretching mode of nickel carbon bond. 
In view of this unmistakable evidence, no other conclusion than that of the 
presence of C-Ni bond could be drawn. The formation of the Ni-C bond at the reaction condition used here is not surprising because of large
surface to volume ratio in the graphene prepared which is therefore highly reactive.   
Figure 3(b) shows the EDX spectra of the system under study. From the EDX 
spectra it can be seen that no magnetic impurity,other than nickel,is present in the material. 
\begin{figure}
\centering
 \includegraphics[width=8.5cm]{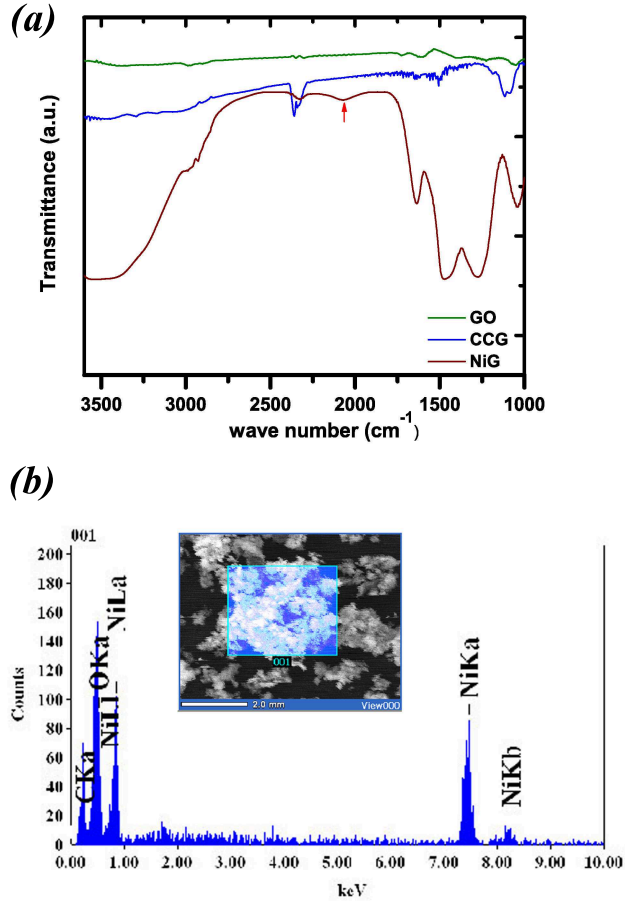}

\caption{(a)FTIR spectrograph of GO,CCG and NIG. (b) EDX spectra of the material under investigation.}
\label{fig.3}
   \end{figure}

 \begin{table}
\small
\caption{Elemental Analysis from EDX spectra of the specimen synthesized.}
\label{tab.1}
\begin{tabular*}{0.5\textwidth}{@{\extracolsep{\fill}}lll}
 \hline
Element & Atomic ${\%}$ \\
 \hline 
Carbon & 42.32 \\
Nickel & 19.07 \\
Oxygen & 38.61 \\
\hline
\end{tabular*}
\end{table}


The elemental analysis as obtained from the
EDX data is shown in table 1.
The amount of oxygen shown in table 1  comes from the $H_{2}O$ molecules present in the system. It may be mentioned here that H can not be
detected by the EDX.
This has been made amply clear by earlier authors who prepared graphene by a similar method of chemically exfoliating GO.\cite{gracnr}
 We have therefore, calculated the atomic percent of nickel after neglecting the amount of oxygen shown in table 1 . 

This suggests that 31 atomic ${\%}$ nickel is present in graphene network.
It is to be noted that oxygen detected by EDX do not form any part of the graphene phase.
 From the selected area electron diffraction (SAED) image [figure 1(c)]
it is concluded that neither nanoparticles of nickel nor any of its oxides have been formed. So, it can be safely concluded that nickel was adsorbed on the surface of
the graphene,during the simultaneous reduction of GO and $Ni^{2+}$ to graphene and atomic nickel respectively. This kind 
of structure has already been studied theoretically,\cite{santosprb,rigoprb} which proves the stability of such structures. 
  \subsection{DC magnetization}
      For non-equilibrium systems, like spin glasses and superparamagnets, history effects in dc magnetization is a generic feature
 i.e. there is irreversibility in the data describing magnetization as a function of temeperature.
 We observe a bifurcation in zero field cooled (ZFC) and field cooled (FC) curves in the magnetization-temperature data [figure 4] for the nickel adsorbed graphene.
The data clearly show a pronounced irreversibility at T $\approx 20$ K. This shows the occurrence of magnetic history effect in the system. 
\begin{figure}
\centering
 \includegraphics[width=8.5 cm]{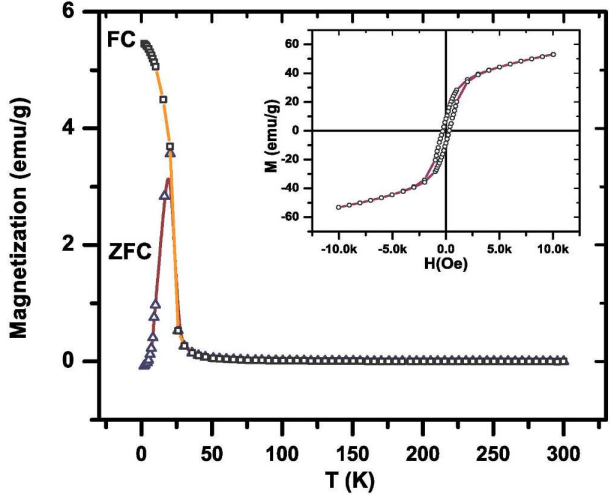}

\caption{ The dc magnetization as measured in field-cooled (FC) and zero-field-cooled (ZFC) conditions at 100 Oe, exhibiting pronounced
irreversibility at the freezing point. The inset shows the non-saturating {\it M-H} isotherm as measured at 10 K.  }
\label{fig.4}
   \end{figure}
In the inset of figure 4, we have shown the M-H isotherm measured at 10 K indicating the presence of a finite coercivity. It is interesting to note
that the deviation of the FC magnetization from the ZFC at the freezing point (temperature at which ZFC peak occurs) is a feature, however not exclusive,
for the canonical spin glass system. This kind of behavior is also expected in the case of superparamagnets having narrow volume distribution. 
In both the cases of superparamagnets and spin glasses, finite dipolar interaction between the spins results in the deviation of FC -ZFC curves at
temperatures lower than blocking or freezing temperatures and FC magnetization increases continuously as the temperature is lowered.\cite{luoprl}
It may be noted here that there have been recent reports on similar behavior of low temperature magnetization peaks obtained in nickel\cite{hejpcm,chenjpcm} and 
 nickel carbide\cite{hejap1,hejap2} nanoparticles respectively. We discuss the implications as follows. In the case of nickel nanoparticles 
a peak is observed but the irreversibility in the magnetization/temperature behavior continues upto room temperature and above. This is the signature of  
superparamagnetism which is consistent with the fact that the systems show room temperature ferromagnetism.On the contrary,in our system, the irreversibility dissappears
after 20K. Also the value of the coercivity becomes zero, indicating that the system is in a paramagnetic state.
Comparing the data of $Ni_{3}C$ nanoparticles\cite{hejap1} with those of our own, we note, that in our case the magnetization value is two orders of magnitude higher. A robust
ferromagnetism is observed below the freezing temperature rather than a weak ferromagnetism observed in the nickel carbide nanoparticle system.\cite{hejap1}
We therefore, rule out the possibility of any  $Ni_{3}C$ nanoparticles being present in our sample. Also the peak in the ac susceptibility versus temperature
plot was reported to be at $\sim 10 K$, whereas, in our system it was found to be around 20K (see next section).
\par
From the above discussion it follows that neither Ni nor  $Ni_{3}C$ nanoparticles are present in the system synthesized by us. Both the compositional analysis (previous section)
and comparison of our magnetic data with carefully conducted magnetization studies on nano nickel and nano  $Ni_{3}C$ done earlier substantiate our description 
of the system under study to be nickel adsorbed graphene.
   \subsection{AC magnetization} 
       In principle, the time dependent susceptibilities might give detailed insight into the dynamics of freezing.
To probe the dynamics of the spin system, we measured the ac susceptibility of the sample at low magnetic field (0.5 Oe). The temperature dependence of
the real $(\chi{'}_{ac})$ and imaginary $(\chi{''}_{ac})$ parts of the ac susceptibility of the present system are shown in figures 5(a) and 5(b) respectively
for frequencies ranging from 10 Hz to 1kHz.It can be seen that the peak height in the $(\chi{''}_{ac}-T)$ is increased and shifted towards high temperature 
side with the increasing frequency.It should be noted that for a conventional ferromagnet the imaginary part of the ac susceptibility
vanishes above and below the peak temperature, but remains non-zero in case of a spin glass system.\cite{binderrmp,ic45} 
\begin{figure}
\centering
 \includegraphics[width=8.5 cm]{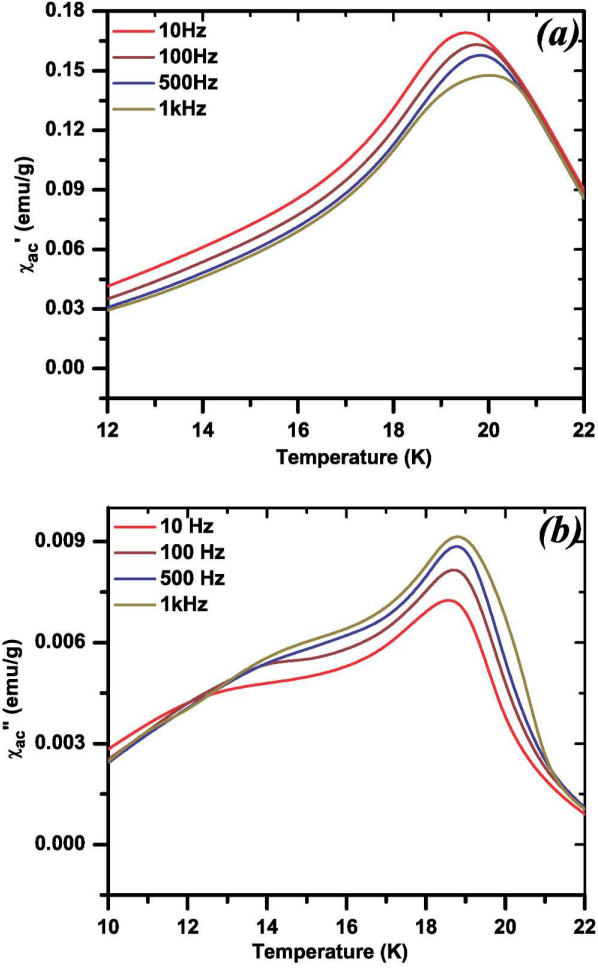}

\caption{ The frequency dependence of the (a) real and (b) imaginary part of the linear susceptibility as measured at 0.5 Oe
with the ac frequency being varied from 10 Hz to 1kHz.}
\label{fig.5}
   \end{figure}
\par
 In the case of a spin glass, both the real and imaginary parts of the ac susceptibility show a frequency dependent cusp or maximum at a temperature, called 
effective spin glass transition temperature $T_{f}(\omega)$, the value of which should increase with an increase in frequency. We observe this particular 
behavior in our system as shown in figure 5. The frequency dependence of the peak temperature may be small, but this kind of small but clear dependance 
was previously observed in case of  canonical spin glass systems.\cite{prl84,mydoshbook}
We have calculated the initial frequency shift of $T_{f}$ from the frequency dependance of the peak temperature, by employing;\cite{prb72,prb74}
\begin{equation}
 \delta T_{f}=[\frac{\Delta T_{f}}{T_{f}\Delta log_{10}\omega}]
\end{equation}
where, $\delta T_{f} $ is the relative change in freezing temperature, $\Delta T_{f}$ is the total change in the $T_{f}$ in the frequency interval, $\Delta log_{10}\omega$
is the frequency interval.In our system $\delta T_{f}$ comes out to be 0.01. 
For canonical spin glasses,the value of $\delta T_{f}$ lies between 0.0045 to 0.06 \cite{mydoshbook,prb72,prb67} e.g.  
CuMn alloy shows a value of 0.005 and AuFe has the value 0.01 \cite{mydoshbook}.
On the other hand for known superparamagnets the value of $\delta T_{f}$ is larger than 0.1.\cite{prb72,mydoshbook}
The ratio of the peak intensity of $Im(\chi_{ac})$ to $Re (\chi_{ac})$ is found to be 0.04, which is of the same order of magnitude as observed previously
by Nishioka et.al. \cite{jpsj69} on their study of canonical spin glass behavior in $Ce_{2}AgIn_{3}$.\\   
A widely used experimental technique, to measure the \textquoteleft characteristic time\textquoteright ( $\tau$ ), for describing the dynamical
fluctuation time scale, is obtained from the observation time.\cite{binderrmp,mkdprb} The characteristic relaxation time of the spin system, at the in-phase
susceptibility maximum corresponds to the observation time t $\sim$ $\frac{1}{f}$.\cite{souletieprb,namprb,sandlundprb} \\
As this frequency dependent maximum of ac susceptibility is not a generic feature of  spin-glass systems only, and can also be found for superparamagnets
it is often difficult to distinguish between spin-glass and superparamagnetic behavior.\cite{binderrmp,mydoshbook}
 
A non-interacting superparamagnetic cluster's relaxation time follows the $N\acute{e}el$-Arrhenius law \cite{neel,souletieprb}
\begin{equation}
 \tau=\tau_{0} exp(\frac{E_{a}}{k_{B}T_{f}})
\end{equation}
where $E_{a}$ is the anisotropy energy barrier, $k_{B}$ is the Boltzman constant, and $T_{f}$ is the peak temperature. $\tau_{0}$ depends on the 
gyromagnetic precession time and is usually of the order of $10^{-13}-10^{-10}$ s. In our analysis $(\tau-T_{f})$ data when fitted with equation (2)
give values $\frac{E_{a}}{k_{B}} = 1250 \pm 320 $ K and $\tau_{0} \sim 10^{-30} (\ll 10^{-13})$ s. These are unphysical quantities and hence we
conclude that observed susceptibility variation is not caused by the non-interacting superparamagnets.
\par 
In glassy systems, the strong temeperature dependence of $\tau$ is frequently described by the law given by Vogel-Fulcher \cite{binderrmp,vogel,fulcher}
\begin{equation}
 \tau=\tau_{0} exp[\frac{E_{a}}{k_{B}(T_{f}-T_{0})}]
\end{equation}
\begin{figure}
\centering
 \includegraphics[width=8.5 cm]{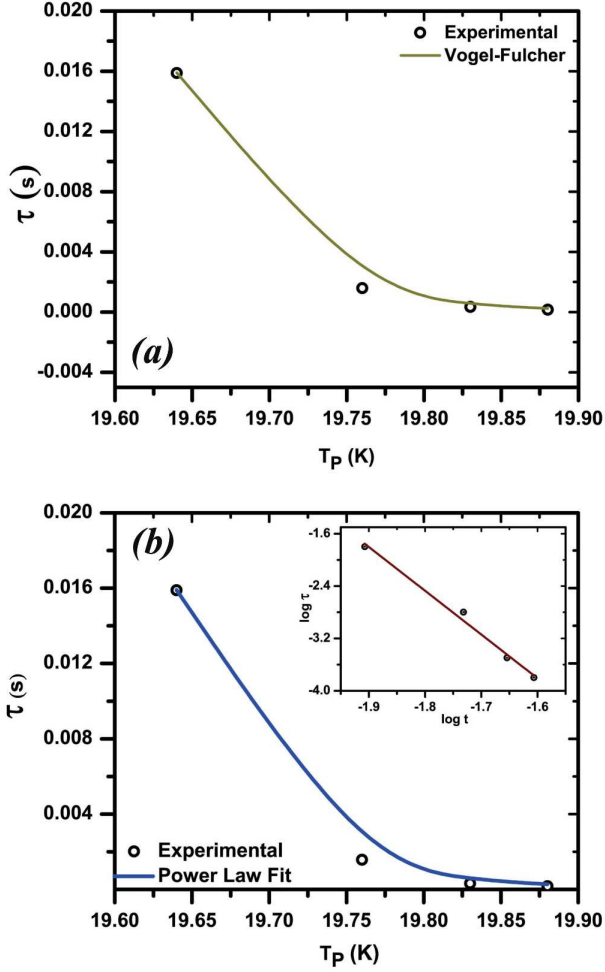}

\caption{ The experimental variation of $\tau$ with peak temperature ({\it${T_{P}}$}) and its fit with (a) Vogel-Fulcher Law and 
(b) Power Law. In the inset of (b) the variation of log $\tau$ with logarithm of reduced temperature ({\it t}) shows a linear behaviour.}
\label{fig.6}
   \end{figure}
 where, the characteristic temperature, $T_{0}$ was introduced in an {\it ad-hoc} manner. The least square fitted results gave values 
$\frac{E_{a}}{k_{B}} = 11.93 \pm 0.56 $ K, $T_{0}=18.92 \pm 0.03 $ K with $\tau_{0}= 9.35 \times 10^{-9} $ s.
The value of $\tau_{0}$ obtained for the system indicates the presence of conventional spin glass. A relatively larger value ($\sim 10^{-7}s$) 
is expected for the interacting magnetic spin clusters.\cite{prl79}
 The data and the theoretically fitted curve have been shown in figure 6(a). 
It should be pointed out that in our system temperature dependence of the susceptibility maximum is small compared to a spin glass system reported earlier.\cite{prl61}
However, the nature of results is identical, the difference being a low value of the activation energyfor spin flipping in our system. Using Vogel-Fulcher analysis,
our results give $\frac{E_{a}}{k_{B}} = 11.93 $ K whereas, Gunnarsson et.al. \cite{prl61} results give $\frac{E_{a}}{k_{B}} = 56.2 $ K.
The value of $T_{0}$ suggests \cite{binderrmp} that there is  RKKY (Ruderman-Kittel-Kasuya-Yosida) type of interaction between the spins of atomic nickel,
mediated via the highly conducting electrons of graphene,operative in our system. There are few theoretical works that describe that the RKKY interaction operative in 
graphene systems with adatoms on it, or having defects \cite{jpcm20,prl91,prb83,prb84} is short ranged because of low density of states of electrons near the Fermi level.
However, in our method of sample preparation we obtained a few layers of graphene (FLG) as evidenced from the high resolution electron micrograph
and x ray diffractrogram [see section 3.1]. In the case of FLG systems the electronic structure will be modified and 
there will be finite density of states at the Fermi energy.\cite{prl96,rmp81,ajp46} This will make the RKKY interaction between the atomic spins of nickel via the electrons
of graphene quite feasible.
\par
The sharp cusp in the temperature dependent ac susceptibility at low magnetic field indicates this to be a case of continuous phase
transition. In case of a continuous phase transition, as the transition temeperature is approached from above, the correlation length ($\xi$) 
diverges as $ \frac{\xi}{a} = t^{-\nu} $ where,  t[=$\frac{T_{f}-T_{g}}{T_{g}}$] is the reduced temperature,with $T_{g}$ as true glass transition
temeperature, $T_{f}$ is freezing temperature, \textquoteleft a\textquoteright being the average distance between the interacting spins, $\nu$ is the critical exponent of
the spin correlation length $\xi$.  If the conventional critical slowing down while approaching $T_{g}$ from higher T side is assumed, the relaxation
time $\tau$ is related to $\xi$ as $\tau \propto (\frac{\xi}{a})^{z}$, where, \textquoteleft z \textquoteright is the dynamic critical exponent.
Thus the temperature dependence of the relaxation time $\tau$ can be expressed by a power law, \cite{binderrmp,nair}
 \begin{equation}
 \tau=\tau_{0} [(\frac{T_{f}}{T_{g}})-1]^{-z\nu}
\end{equation}  
Figure 6(b) shows the fitting of the experimental data with the above mentioned power law, with $\tau_{0}$ as the microscopic flipping time of the
fluctuating entities. The comprehensive fitting gives $z\nu = 5.9 \pm 0.10$, $\tau_{0} = 10^{-13}$s. and $T_{g}= 19.5 \pm 0.02$ K. The inset of the
figure 6(b) shows the variation of the relaxation time with reduced temperature (both in log scale) and the linear fit of the data. The linear fit also 
gave the same order of magnitude value of the parameters as previously obtained. 
The value of the critical exponent ($z\nu$) is characteristic of a spin glass system ($5 \leq z\nu \leq 11$ )\cite{mydoshbook} and quite different from those characteristic
of regular ferromagnets ($1.2 \leq z\nu \leq 2$ ). \cite{chaikinbook}
It should be mentioned here that for all the analysis of the relaxation time, we have considered $\tau=(2 \pi f)^{-1}$ and while deducing
$T_{f}$ the maximum of  $(\chi{'}_{ac})$ was used. Alternatively, the temeperature derivative of real or imaginary parts of ac susceptibility
could be used. In case of  $\chi{'}$ or  $\chi{''}$ the prominent peak was observed at the same temperature, as expected, but in ($\chi{''}_{ac}$-T) data
we observe another hump at the lower temeperature side which is surprisingly absent in ($\chi{'}_{ac}$-T) data and it needs clarification. We have tried to 
analyse them with previously mentioned models, viz; $N\acute{e}el$-Arrhenius; and Vogel-Fulcher; but none of them was able to explain their occurrence,
giving unphysical values of the parameters with huge errors. However, the prominent peaks in the temperature dependence of the real and imaginary 
parts of the ac susceptibility in the light of a relaxation mechanism have been successfully analysed. \\
From the dynamical scaling analysis of the ($\tau$-T) we found the microscopic flipping time to be of the order of $10^{-13}$ s, which is of the same order as that found 
in conventional spin glasses.\cite{prl61} If it were a cluster like spin glass then the $\tau_{0}$ value should have been of the order of $10^{-10}s$ as reported 
previously by different workers on their studies on different cluster like spin glass systems.\cite{mkdprb,jap95}
This is quite easy to understand that the microscopic magnetic entities are atomic spins of nickel and not nanosized clusters of ferromagnetically coupled spins.
 The present analysis gives the value of the critical exponent $z\nu$ as 5.9, which is not only close to the theoretically predicted
value for the short range Ising and Heisenberg spin glasses,\cite{fisherbook} but is also the experimentally observed value for well known conventional spin glass systems like
CuMn alloy at 4.6 at.${\%}$.\cite{souletieprb}. \\
From the analysis, it is evident that both the Vogel-Fulcher and the power law successfully explain the relaxation mechanism, with different sets of $T_{0}$,
$T_{g}$ and $\tau_{0}$ values. The difference between $T_{0}$ and $T_{g}$ is quite small $(\sim 0.58 K)$, whereas, $(\tau_{0})_{PL}$ differs from  $(\tau_{0})_{VF}$
by 4 orders of magnitude. This kind of different $\tau_{0}$ values, using  power law and Vogel-Fulcher law was previously obtained by Souletie and Tholence \cite{souletieprb}
from their study on different spin glasses.
\par
We have determined the spin glass order parameter (q), starting from the susceptibility using the following equation\cite{binderrmp}
\begin{equation}
 \chi(T)=\frac{C(T)[1-q(T)]}{T-\theta(T)[1-q(T)]} 
\end{equation}
where, both C(T) and $\theta(T)$ are temperature independent constant in mean field theory, but varies slowly near the spin glass
transition temperature.To evaluate q(T), $[(\chi{'}_{ac})]^{-1}$ was extrapolated and from that $\theta(T_{f})$ and
 $C(T_{f})$ were estimated. Using Eq.(5) q(T) was found to be \cite{prl72}
\begin{figure}
\centering
 \includegraphics[width=8.5 cm]{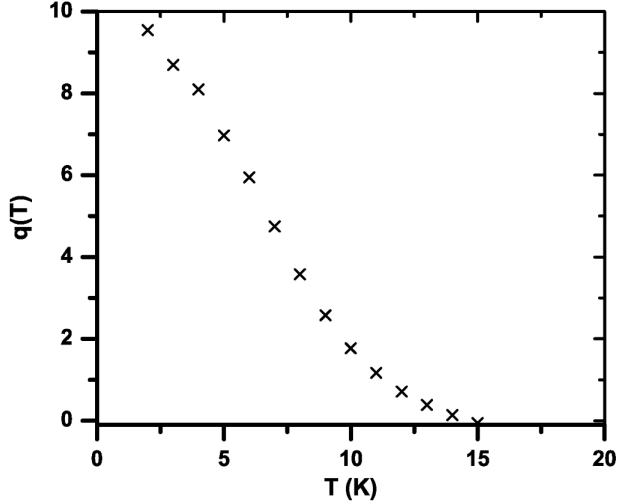}

\caption{ Spin glass order parameter,q(T), derived from the data of inverse susceptibility using Eq.(5).}
\label{fig.7}
   \end{figure}


\begin{equation}
 q(T) \approx \frac{1-T\chi(T)}{[C(T_{f})+\theta(T_{f})\chi(T)]} 
\end{equation}
The values of q(T) as derived by Eq.(6) is shown in figure 7. it is observed that the measured q(T) decreases almost linearly with increasing temperature.
 The analysis of the data using $T_{f}=19.48 K$ shows a power law for q(T), as predicted by the mean field theory.\cite{binderrmp}
\par
\subsection{Thermoremanent magnetization}
\begin{figure}
\centering
 \includegraphics[width=8.5 cm]{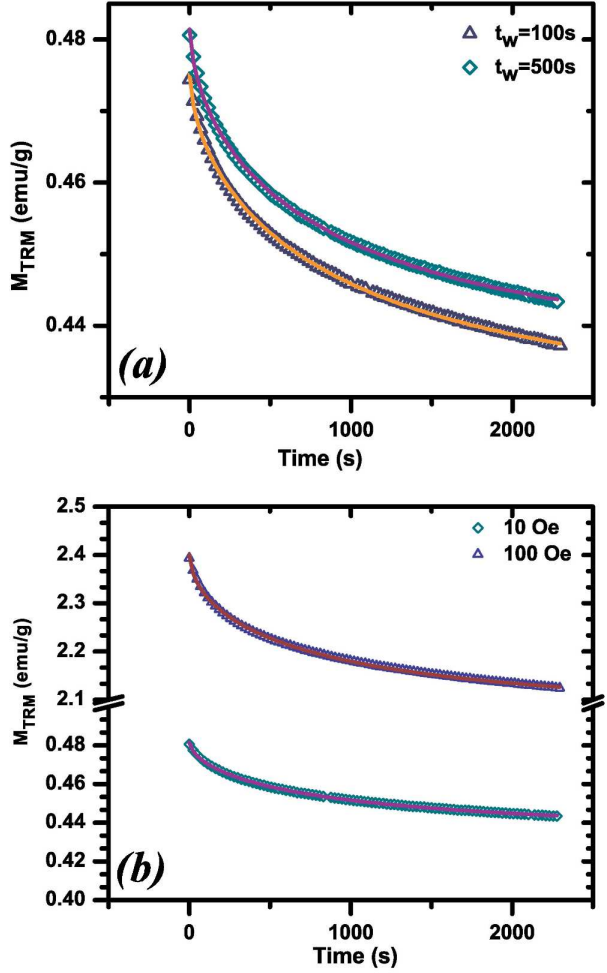}

\caption{ (a)Time decay of {\it $M_{TRM}$} as measured at 0.72 $T_{f}$ after cooling in the presence of a field of 10 Oe and  for different wait times ($t_{w}$).
(b) The variation of $M_{TRM}$ with time for different magnetic field shown in a same graph.
The solid lines are the theoretically fitted curves with the stretched exponential function [eq.7].  } 
\label{fig.8}
   \end{figure}

One of the important characteristic features of spin-glasses is the phenomenon of aging. This occurs due to the breakdown of time translational invariance
(as the magnetization evolves with time) of the response of the system under external perturbation. In spin glasses, a common way to explore aging is the
decay of thermoremanent magnetization $(M_{TRM})$ with time. As the behavior of spin glass is complicated below freezing temperature due to ageing, we have taken an particular 
approach to measure the TRM relaxation.
 The TRM measurements were done at $0.72 T_{f}$ (14 K) after cooling the sample under a
magnetic field (10 Oe) from $2.0 T_{f}$ (40 K). After stabilizing the temperature, the system was made to wait for a time $t_{w} $. Then the magnetic field was cutted off
 and the evolution of magnetization with time was recorded. 
Out of various functional forms, to describe the time variation of remanent magnetization for spin glass systems,we have adopted 
one of the most commonly used relations, viz; a stretched exponential function \cite{mkdprb}
\begin{equation}
 M(t)=M_{0} + M_{r} exp[-(\frac{t}{\tau_{r}})^{\beta}]
\end{equation}

where, $M_{0}$ relates to an intrinsic ferromagnetic component, and $M_{r}$ to a glassy component. the time constant $\tau_{r}$ depends on T and $t_{w}$, whereas,
$\beta$ is a function of T. For $\beta = 1$, the relaxation involves
the activation against single energy barrier, and for $0 < \beta < 1$, it stands for SG systems; and $\beta =0$ implies there is no relaxation at all.
Our (TRM-T) data fitted well with the above mentioned stretched exponential function, and did not follow the simple logarithmic decay. From the fitting the value
of $\beta$ was extracted as $0.61 \pm 0.007$. Both the experimental and the theoretically fitted curves have been shown in figure 8. Figure 8(a) shows the evolution of 
thermoremanant magnetization for different wait times (viz; $t_{w} = 100 s $ and $ 500 s $ ) cooled under a 
magnetic field of 10 Oe. It is seen that the $M_{TRM}$ evolved in identical manner, and relaxes more slowly for longer wait time. In figure 8(b) we presented the magnetic
field dependence of M(t). These showed that aging has an effect on the magnetic relaxation in the system under study as it was observed for different spin glass 
like systems.\cite{prb67,prb63,prb44,prb30,epl2,prb33}


\section{Conclusions}
In summary, we have synthesized nickel adsorbed graphene via simultaneous reduction of GO and nickel ions. 
 The dc magnetization study along with frequency dependent ac magnetization and magnetic relaxation studies confirmed the spin glass like
behavior in the system. The observed frequency dependence of the peak temperature in the ac susceptibility curve was sucessfully analysed by using both the
Vogel-Fulcher and power law.The dynamic scaling analysis gave the values of the critical exponent $z\nu=5.9 \pm 0.10$ and microscopic flipping time $\tau_{0} \sim 10^{-13}s$.
The value of $\tau_{0}$ indicates the presence of conventional spin glass in the system under study. This nickel adsorbed graphene system has a true spin glass transition
 temperature $T_{g} = 19.4 \pm 0.02$ K as obtained from the dynamical scaling analysis. The thermoremanent magnetization decay was successfully explained by the stretched
 exponential function, giving the value of the exponent as $0.61 \pm 0.007$. The value of the interaction parameter $T_{0}$ from Vogel-Fulcher analysis suggests 
an RKKY type of interaction between the atomic spins of nickel,where the interaction is mediated via the conducting electrons of graphenic carbon network.  


\section*{Acknowledgements}
Authors thankfully acknowledge the Department of Science and Technology, Govt.of India, New Delhi for support under an Indo-Australian Project on Nanocomposites. 
Sreemanta Mitra thanks University Grants Commission, New Delhi,India
for awarding a Senior Research Fellowship. DC thanks Indian National Science Academy, New Delhi,India for an Honorary Scientist's position.

\end{document}